# Passive Bias-Free Nonreciprocal Metasurfaces Based on Nonlinear Quasi-Bound States in the Continuum

Michele Cotrufo[1], Andrea Cordaro[2,3], Dimitrios L. Sounas[4], Albert Polman[3] and Andrea Alù[*1,5]

[1]Photonics Initiative, Advanced Science Research Center, City University of New York, New York, NY 10031, USA

[2]Van der Waals-Zeeman Institute, Institute of Physics, University of Amsterdam Science Park 904, 1098 XH Amsterdam, The Netherlands

[3]Center for Nanophotonics, AMOLF, Science Park 104, 1098 XG Amsterdam, The Netherlands

[4]Department of Electrical and Computer Engineering, Wayne State University, Detroit, Michigan 48202, USA

[5]Physics Program, Graduate Center of the City University of New York, New York, NY 10016, USA

*Nonreciprocal devices – in which light is transmitted with different efficiencies along opposite directions – are key technologies for modern photonic applications, yet their compact and miniaturized implementation remains an open challenge. Among different avenues, nonlinearity-induced nonreciprocity has attracted significant attention due to the absence of external bias and integrability within conventional material platforms. So far, nonlinearity-induced nonreciprocity has been demonstrated only in guided platforms using high-Q resonators. Here, we demonstrate ultrathin optical metasurfaces with large nonreciprocal response for free-space radiation based on silicon third-order nonlinearities. Our metasurfaces combine an out-of-plane asymmetry – necessary to obtain nonreciprocity – with in-plane broken symmetry, which finely tunes the radiative linewidth of quasi-bound states in the continuum (q-BICs). Third-order nonlinearities naturally occurring in silicon, engaged by q-BICs, are shown to enable over 10 dB of nonreciprocal transmission while maintaining less than 3 dB in insertion loss. The demonstrated devices merge the field of nonreciprocity with ultrathin metasurface technologies, offering an exciting functionality for signal processing and routing, communications, and protection of high-power laser cavities.*

Nonreciprocal electromagnetic devices transmit light asymmetrically along opposite directions, forming key components to achieve ultimate control over the flow of light. However, nonreciprocal transmission is hard to achieve in conventional media: Lorentz reciprocity [1] dictates that, in any system with permittivity and permeability tensors that are *symmetric*, *time-invariant* and *linear*, the transmission between a source and a detector is invariant if these are swapped. Breaking reciprocity, thus, requires lifting at least one of these conditions. Standard approaches for light isolation involve applying a dc magnetic bias to magneto-optical materials, which makes the permittivity tensor asymmetric. More recently, nonreciprocity has been achieved with time-variant materials, whereby some material properties, such as the refractive index, are modulated in time [2]–[11]. Finally, reciprocity can be broken by exploiting electromagnetic nonlinearities [12]–[20]. This approach has recently received significant attention, due to the absence of any external form of bias and the universal working principle, directly integrable in a variety of conventional photonic platforms [21]. When an electromagnetic resonator couples asymmetrically to two input/output ports, the same power injected from different ports gives rise to different intra-cavity field intensities. In linear systems, such internal asymmetry is not sufficient to break reciprocity, and the port-to-port transmission remains the same in both directions. However, if the resonator is filled with a material with nonlinear response, such as an intensity-dependent permittivity, different intracavity intensities create different permittivity profiles, enabling large asymmetries in the power flow for opposite directions [21]. Remarkably, this mechanism does not require any applied bias – in essence it is the signal itself to self-bias the device – and it does not require any absorption, because the unwanted beam is reflected rather than being absorbed as required in the case of magnetic-based isolators. While general constraints based on passivity and time-reversal symmetry prevent these devices from working as conventional isolators under simultaneous two-port excitation [22], they constitute an appealing technology for applications such as nonreciprocal routing of pulsed signals [20] and protection of high-power lasers. Indeed, due to its simplicity and general applicability, the nonlinearity-based route to nonreciprocity has been successfully demonstrated in various frameworks, such as integrated Si and InP micro-cavities operating in the near-infrared [16],[20], microwave circuits [19], and atomic systems [22]–[24]. All the devices investigated so far involve integrated systems coupled to optical waveguides or transmission lines, since in these devices wave-matter interactions can be carefully controlled and enhanced, and the typically weak optical nonlinearities can be engaged in a controllable fashion. A few theoretical proposals [17], [18], [23]–[25] have suggested that these phenomena may be also translated to optical metasurfaces coupled to propagating free-space plane waves, which may be used to realize free-space fully-passive flat nonreciprocal devices. However, the experimental demonstration of nonlinearity-induced nonreciprocity in optical metasurfaces has been so far elusive, mainly due to the weak nonlinearities of the involved materials, and the corresponding stringent requirements in terms of operating intensities and low material loss.

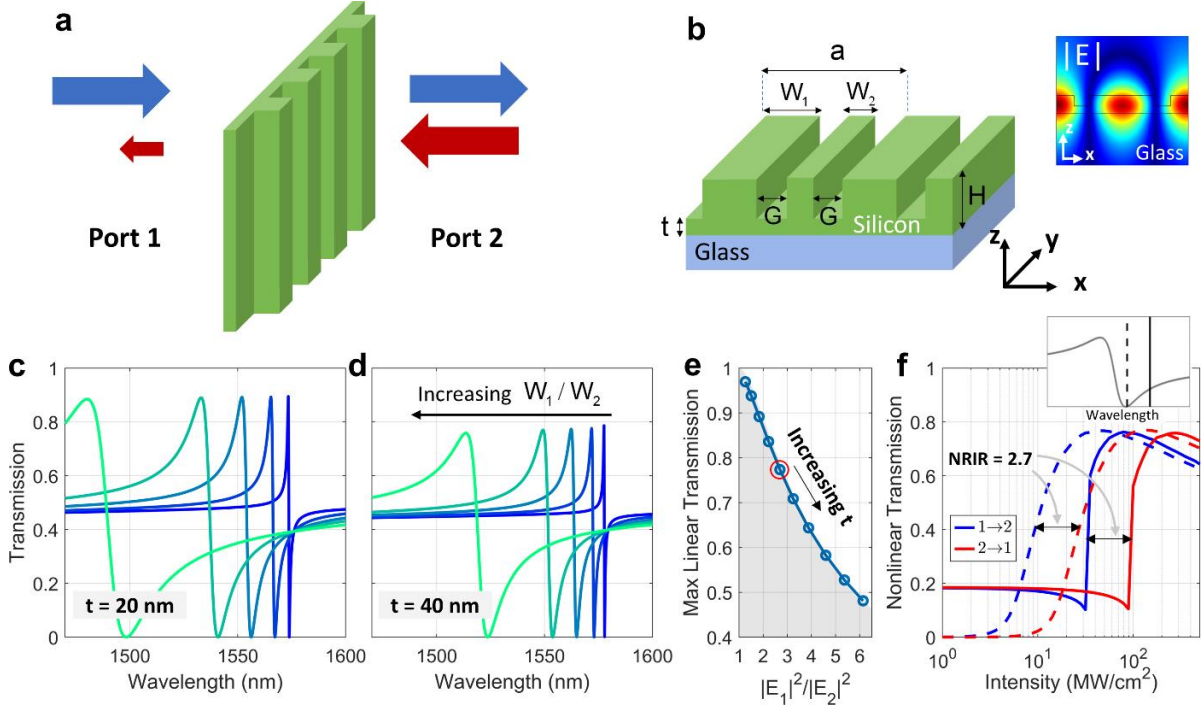

**Figure 1**. (a) Schematic of the nonreciprocal metasurface: by combining structural asymmetry and material nonlinearity, a monochromatic beam impinging from either of the two sides of the device experiences markedly different transmission levels. (b) Geometry and design parameters (additional details in text). Inset: field profiled of the q-BIC excited by incoming plane waves. (c-d) Calculated transmission spectra for $t = 20$ nm (panel c) and $t = 40$ nm (panel d), for fixed values of lattice constant ($a = 750$nm), gaps ($G = 100$ nm) and total thickness (H = 100 nm), and different in-plane asymmetry $W_1/W_2$ ranging from 1.2 (dark blue lines) to 3.4 (light green lines). (e) Scatter plot showing, for devices with fixed value of $W_1/W_2=2.14$, the maximum linear transmission and field asymmetry $|E_1|^2/|E_2|^2$ for different values of $t$. (f) Calculated nonlinear transmission for the device marked by the red circle in panel e, for excitation from port 1 (blue lines) and port 2 (red lines), and for two excitation wavelengths, indicated by solid and dashed lines in the top-right inset.

In this work, we experimentally demonstrate the emergence of strong nonlinearity-induced nonreciprocity in amorphous silicon metasurfaces coupled to free-space radiation in the near-infrared regime (Fig. 1a). In order to create an asymmetric coupling between the metasurface and plane waves propagating along the two normal directions, the out-of-plane symmetry of the device is broken by leaving a thin unpatterned layer, whose thickness can be controlled to maximize nonreciprocity when combined with the third-order nonlinearities naturally occurring in silicon [18]. A major challenge to enable large nonreciprocity in metasurfaces is the typically weak interactions of light with ultrathin devices, which, combined with the poor nonlinearity of optical materials, results in negligible transmission asymmetries, and explains the lack of an experimental demonstration of these concepts to date. Here, we address this issue by introducing tailored in-plane broken symmetries that carefully control a quasi-bound state in the continuum (q-BIC) [26]. In turn, the q-BIC linewidth tailors the metasurface resonant response, enhancing the nonlinear interactions with the incoming waves and hence minimizing the operating intensity, while at the same time keeping a large transmission contrast. Based on these principles, we are able to demonstrate transmission contrasts larger than 10 dB and insertion loss smaller than 3 dB for peak intensities smaller than 50 MW/cm$^2$, only limited by the material nonlinearity strength. Moreover, we experimentally demonstrate that the range of intensities over which nonreciprocity occurs can be fully controlled by the vertical asymmetry of the metasurface. In agreement with recent theoretical results [18], we also experimentally demonstrate a trade-off between the extent of input intensity range over which nonreciprocity occurs and the minimum insertion loss. Tailoring the metasurface geometry, we are able to operate close to the bounds allowed by this trade-off.

## Results

**Device Design and Numerical Optimization.** Figure 1(b) shows the geometry of our metasurface, consisting of a 1D amorphous silicon grating of total thickness $H = 100$ nm placed on a glass-like substrate to allow mechanical handling. The grating is uniform along the *y* direction and periodic along the *x* direction with period *a*. The two excitation/collection ports (denoted 'Port 1' and 'Port 2' in Fig. 1a) correspond to plane waves propagating in opposite directions normal to the metasurface plane (*z*-direction) with impinging electric field polarized along the direction of the nanowires forming the metasurface. Nonreciprocity is achieved if the excitations coming from the opposite ports couple with different efficiencies to the same optical mode supported by the metasurface. In other words, the same intensity, injected from opposite directions, must result in different steady-state intracavity field intensities of the optical mode (see also inset of Fig. 1b). In order to induce and control such *electromagnetic asymmetry*, a residual silicon layer of thickness $t < H$ is left unpatterned. A device with $t = 0$ is symmetric along *z* (apart from the small asymmetry induced by the substrate), and thus it is expected to provide reciprocal wave transmission at any input power, independent of the nonlinearity.

In order to enable strong nonlinearity-induced nonreciprocity, electromagnetic asymmetry is not sufficient, and strong nonlinear interactions are vital. We maximize these phenomena by carefully controlling the radiative linewidth of the targeted resonant mode. Indeed, on one hand it is desirable to reduce the radiative linewidth and hence maximize the Q-factor of the metasurface to strengthen the nonlinear interactions and minimize the required intensity to trigger these nonlinear phenomena [21]; on the other hand, in order to observe a large transmission contrast the resonance linewidth must be larger than the linewidth of the impinging laser, which is particularly important when using pico- or femto-second pulsed lasers. Too narrowband responses also typically imply very selective angular responses, which hinder the possibility of focusing light on the sample to enhance the input intensity. In our device, we address these trade-offs by precisely controlling the Q-factor and frequency response of the metasurface through q-BIC engineering [26]. Q-BICs have been proven to be a very useful platform to boost the Q-factor of metasurfaces and enhance nonlinear phenomena, such as lasing and second harmonic generation [27]–[31]. Here, we show that q-BICs, combined with nonlinear responses, can be used to dramatically enhance nonreciprocal wave transmission, by enhancing the metasurface Q factor while at the same time maintaining a large contrast in the Fano-like transmission spectra lineshape. We consider a unit cell composed of two silicon wires of lateral widths $W_1$ and $W_2$, separated by even gaps of width $G$ (Fig. 1b). When the unit cell is symmetric ($W_1 = W_2$), the metasurface supports a localized mode that does not couple to free-space radiation due to symmetry, realizing a symmetry-protected BIC. Breaking the in-plane symmetry of the unit-cell ($W_1 \neq W_2$) turns the BIC into a q-BIC with finite radiative decay rate. This leads to the appearance of a Fano profile in the transmission spectrum, whose linewidth is carefully controlled by the unit cell asymmetry. Figures 1(c-d) show the numerically calculated linear transmission spectra of devices with $a$ = 750 nm, $G$ = 100 nm and different values of in-plane asymmetry $W_1 / W_2$ ranging from 1.2 to 3.4 (see horizontal arrow in Fig. 1d), and for $t$ = 20 nm (Fig. 1c) and $t$ = 40 nm (Fig. 1d). These results confirm that the linewidth of the Fano resonance can be continuously tuned by controlling the value of $W_1 / W_2$. The inset of Fig. 1b shows the electric field intensity profile induced in a representative device by a plane-wave excitation resonant with the Fano transmission minimum. To quantify the electromagnetic asymmetry, we define [18] the ratio $\kappa = |E_1|^2/|E_2|^2$ of field intensities $|E_i|^2$ ($i = 1,2$) induced in the device when the same power is injected from either port 1 or port 2. Devices with larger $\kappa$ feature larger electromagnetic asymmetries, hence leading to larger nonreciprocity when nonlinearities kick in. A second important metric is the maximum linear transmission $T_{\max}$, obtained at the peak of the Fano lineshape, which determines the forward transmission of the device. As noticeable in Figures 1(c-d), $T_{\max}$ is affected by the vertical asymmetry: devices with smaller asymmetries, i.e., smaller values of $t$ for given $H$ (Fig. 1c), feature higher transmission peaks compared to devices with larger asymmetries (Fig. 1d). This trend is not accidental, but instead rooted in a fundamental bound imposed by time-reversal symmetry [32], [18], [33]: for any two-port lossless device supporting a single resonant mode, the maximum transmission is bounded by $T_{\max} \leq 4\kappa/(\kappa^2 +1)$, and $T_{\max} = 1$ can be obtained only with symmetric devices, i.e., $\kappa = 1$. The parameter space allowed by this bound is depicted by the shaded area in Fig. 1e, while the blue circles represent the [$\kappa$, $T_{\max}$] coordinates of different simulated devices with fixed in-plane geometry and different thickness $t$ ranging from 0 to 90 nm. All the considered devices are optimized to maximize the trade-off $T_{\max}$ vs $\kappa$, i.e., they feature the largest admissible $T_{\max}$ for a given electromagnetic asymmetry.

In order to numerically confirm the nonreciprocal response, we first assume that the dominant nonlinearity in silicon is an instantaneous Kerr-like effect, such that the silicon permittivity depends on the local electric field intensity $|\mathbf{E}|^2$ as $\varepsilon_{Si} = \varepsilon_{Si,lin} + \chi^{(3)} |\mathbf{E}|^2$, with $\chi^{(3)} = 2.8 \cdot 10^{-18} m^2/V^2$ [34]. Consider for instance a metasurface with $t = 40$ nm and $W_1 = 375$ nm, $W_2 = 175$ nm, i.e., the device marked by the red circle in Fig. 1e. Figure 1f shows the simulated intensity-dependent port-to-port transmission for excitation from port 1 (blue curves) and port 2 (red curves) for two different excitation wavelengths. When the system is excited by a monochromatic wave tuned to the transmission minima of the Fano lineshape (dashed vertical line in the inset of Fig. 1f), for both impinging directions the transmission is zero at low intensities and it smoothly increases for higher intensities (dashed blue and red curves). However, due to the asymmetric response, the transmission growth with intensity is different for opposite excitation directions, resulting in large nonreciprocal transmission within a range of input intensities. When the excitation wavelength is instead red-tuned with respect to the minima of the linear transmission spectrum (solid lines in Fig. 1f), sharper transitions between low and high transmission values are observed, based bistability [21]. Since for our device the two transmission curves are scaled versions of each other, following [18] we can define the *nonreciprocal intensity range* (NRIR) as the ratio between the intensities $I_1$ and $I_2$, which, when injected from opposite directions, lead to the same level of transmission (NRIR ≈ 2.7 for the device in Fig. 1f). It can be shown that the NRIR is always equal to the linear electromagnetic asymmetry, $\text{NRIR} = \kappa = |E_1|^2/|E_2|^2$ [18], and thus it does not depend on the excitation wavelength, as also confirmed by Fig. 1f. For any excitation wavelength and propagation direction, the maximum transmission level in the nonlinear scenario (Fig. 1f) is identical to the one obtained in the linear transmission spectra (Fig. 1d). In particular, the aforementioned trade-off between maximum linear transmission and electromagnetic asymmetry leads to a corresponding trade-off for the maximum nonlinear forward transmission and the NRIR, given by [18]

$$T_{max}^{nonliner} \leq \frac{4 \times \text{NRIR}}{\text{NRIR}^2 + 1}. \tag{1}$$

Thus, while a wider NRIR can be obtained by increasing the electromagnetic asymmetry of the metasurface, i.e., by increasing *t*, this comes at the cost of reduced transmission, implying larger insertion loss.

Before discussing the experimental results, we emphasize the different roles played in our design by the *vertical ( $t > 0$ )* and *in-plane ( $W_1 \neq W_2$ ) asymmetries*. The vertical asymmetry is required to ensure that the coupling rates between the optical mode supported by the grating and plane waves propagating along the *+z* and *-z* directions are different; this asymmetry, combined with the material nonlinearity, gives rise to the nonreciprocal behavior. The value of *t* also controls the NRIR and, following Eq. (1), it therefore determines the minimum insertion loss of the device. The in-plane asymmetry, instead, is used as a knob to finely tune the Q-factor of the metasurface and hence maximize the nonlinear interactions, but it is not fundamentally needed to achieve nonreciprocity.

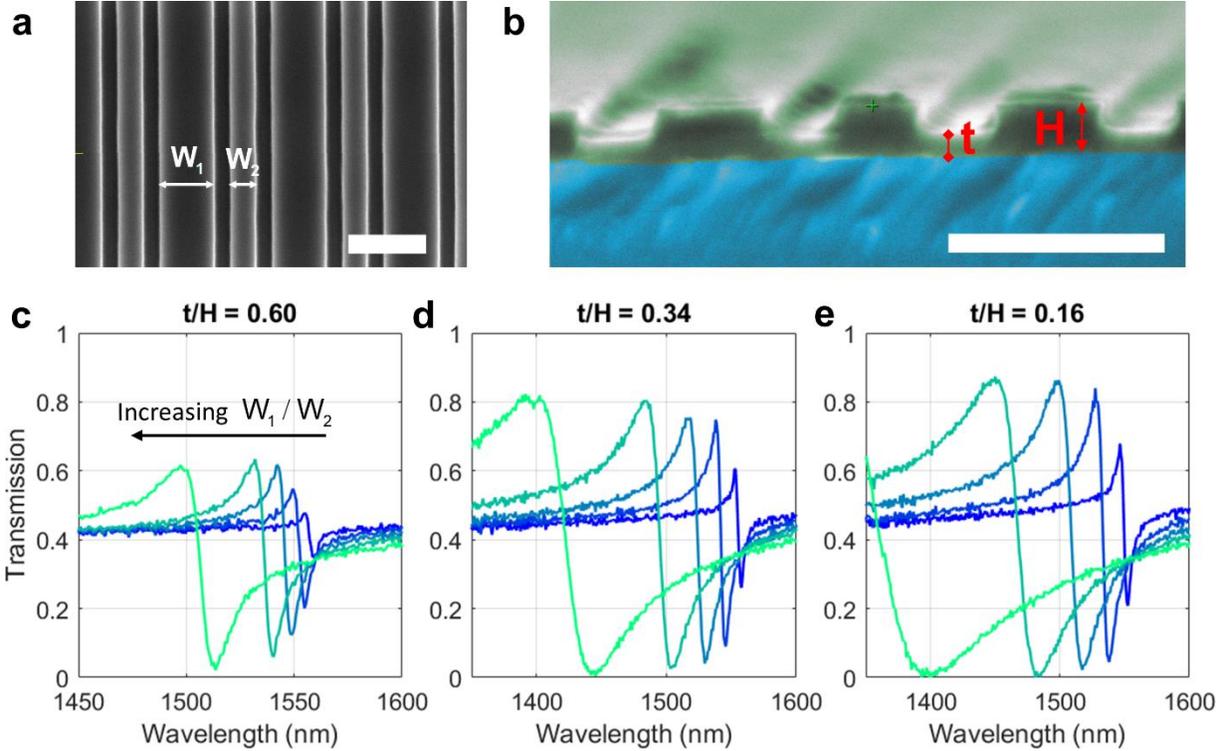

**Figure 2**. (a-b) Top-view (panel a) and false-colored cross-sectional (panel b) SEM micrographs of two fabricated metasurfaces. Scalebars = 500 nm. (c-e). Measured normal-incidence transmission spectra of 15 different devices with fixed lattice constant a = 750nm, gaps G = 100 nm and H = 97 nm. In each panel, the devices have the same residual thickness ($t$ = 58 nm in panel c, $t$ = 33 nm in panel d, and $t$ = 16 nm in panel e), while different colors denote different in-plane asymmetries $W_1/W_2$, ranging from 1.2 to 3.4.

**Linear Characterization**. We fabricated several metasurfaces with geometrical parameters similar to the one used in Fig. 1 (see [35] for details on the fabrication process), and with different values of in-plane asymmetry $W_1/W_2$ and vertical asymmetry $t$. The lattice constant $a$ was varied between 700 nm and 800 nm in order to obtain resonant modes in the 1450 nm – 1650 nm spectral window. Figure 2(a-b) shows top-view and cross-sectional SEM micrographs of two representative devices. A custom-built setup, described in detail in [28], is used to measure the linear and nonlinear response of the devices. The linear transmission spectra were measured by illuminating the sample with linearly polarized broadband light while acquiring the transmitted signal with a spectrometer. Figures 2c shows the transmission spectra of five different devices with $t/H=0.60$ and different values of in-plane asymmetry $W_1/W_2$. In agreement with the simulations in Figs. 1(c-d), each spectrum features a Fano profile whose linewidth grows as $W_1/W_2$ increases. While in simulations [Figs. 1(c-d)] the maximum and minimum transmission do not depend on $W_1/W_2$, in the measured spectra the transmission contrast is reduced as the linewidth gets smaller, e.g., blue curves in Figs. 2c. This is primarily due to unwanted loss and imperfections introduced by the fabrication process, and also by the finite angular and spectral resolution of our setup. As explained above, the maximum transmission level $T_{max}$ is also fundamentally limited by the value of the vertical asymmetry $t$. By reducing the vertical asymmetry, i.e., reducing $t$ for fixed $H$, $T_{max}$ is expected to increase, as confirmed by the measured spectra in Fig. 2d ($t/H=0.34$) and Fig. 2e ($t/H=0.16$).

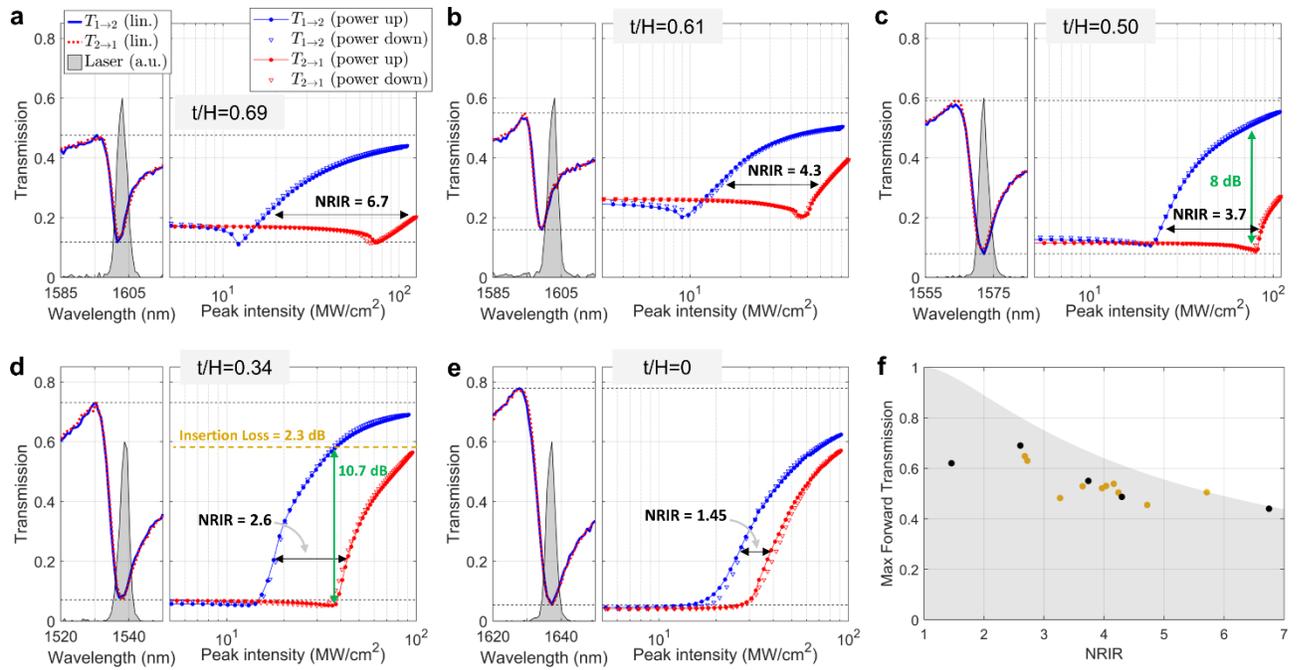

**Figure 3**. (a-e) Linear and nonlinear characterization of five devices with different values of t/H, as indicated by the text in each panel. In each panel, the left plot shows the laser spectrum (shaded gray area) used in the nonlinear measurement, and the linear transmission spectra of the device measured along the two directions (solid blue and dashed red lines), acquired immediately before the corresponding nonlinear measurement. The right plot of each panel shows the measured nonlinear transmission versus impinging intensity, for a wave impinging from port 1 (blue symbols) and port 2 (red symbols), and for increasing and decreasing intensities. (f) Scatter plot showing the NRIR and the maximum forward transmission of the devices in panels a-e (black dots), and several other measured devices not shown here (yellow dots). The grey area shows the trade-off corresponding to Eq. 1.

**Demonstration of Nonlinearity-Induced Nonreciprocity.** To experimentally demonstrate strong nonlinearity-induced nonreciprocal transmission through our metasurfaces, we measured the intensity-dependent transmission of several devices along the two directions. In our setup (see description in [35]), a pulsed laser was weakly focused on the device under test. We first acquired the intensity-dependent transmission along one direction by sequentially sweeping the impinging power up and down, and then we flipped the sample in order to measure the transmission along the opposite direction. In doing so, particular care [35] was taken to make sure that in both measurements the excitation beam impinged on the sample at normal incidence (to avoid spurious shifts of the transmission spectra) and with the same spot size (to ensure that the intensity level was the same), and that the laser impinged at the same position within the grating (to minimize the impact of fabrication-induced inhomogeneities).

Figure 3(a-e) shows the intensity-dependent transmission of five representative metasurfaces with vertical asymmetries ranging from large values (t/H=0.69, Fig. 3a, and t/H=0.61, Fig. 3b) to intermediate values (t/H=0.50, Fig. 3c, and t/H=0.34, Fig. 3d), and to a nominally symmetric device ($t/H=0$, Fig. 3e). In each panel, the left plot shows the linear transmission spectra measured in the two directions (solid blue and dashed red lines) along with the spectrum of the laser (grey shaded area) used in the corresponding nonlinear measurements, while the right plot shows the intensity-dependent transmission curves when exciting from port 1 (blue symbols) and port 2 (red symbols). For each experimental run, we acquired the transmission while sweeping the power up (circles joined by solid lines) and down (triangles). The good agreement between the nonlinear transmission curves recorded for increasing and decreasing power levels confirms that the observed changes in transmission are indeed due to a reversible intensity-dependent shift of the permittivity, and not to any irreversible damage to the device. As anticipated, the device with the largest vertical asymmetry (Fig. 3a, t/H=0.69) also features the widest nonreciprocal intensity range (NRIR = 6.7), while its maximum transmission in the forward direction is limited to ~ 0.44. Importantly, the minimum and maximum transmission levels obtained in the nonlinear curves (horizontal dashed lines in Fig. 3a) match the minimum and maximum transmission levels observed in the corresponding linear transmission spectrum. When reducing the vertical asymmetry (Figs. 3b-d), the maximum forward transmission (blue symbols) increases, while the NRIR progressively shrinks. In the limit of a nominally symmetric device (t=0, Fig. 3e), the range of intensities over which nonreciprocity can be observed is strongly reduced (NRIR ≈ 1.45). The small yet nonzero NRIR observed for *t* = 0 is due to a small residual electromagnetic asymmetry introduced by the glass substrate and by the tilted sidewalls of the grating, as confirmed by numerical simulations [35].

The scatter plot in Fig. 3f shows the NRIR and the maximum forward transmission of the devices in Figs. 3(a-e) (black dots) and of several other measured devices (yellow dots), experimentally confirming the trade-off dictated by Eq. (1). Our devices lie very close to the edge of the trade-off curve, indicating that the maximum transmission is only limited by small intrinsic loss. Moreover, we note that, for some of the devices in Figs. 3(a-e), the maximum input power available in our experiment limits the achievable forward transmission, and thus some of the points in Fig. 3f are actually closer to the boundary of the trade-off region. Remarkably, our devices can achieve large nonreciprocal ratios (defined as the ratio between transmissions in the two directions for the same input intensity), while simultaneously providing low insertion loss. Specifically, for intermediate values of NRIR, such as the device in Fig. 3d (NRIR = 2.6), a nonreciprocal ratio larger than 10 dB is obtained for intensities of about 40 MW/cm$^2$ (green vertical line in Fig. 3d) accompanied by insertion loss of 2.3 dB. Importantly, the values of insertion loss obtained here are close to the minimum value imposed by the fundamental trade-off in Eq. 1 (see also shaded area in Fig. 3f), which affects any single-resonator device. As discussed by recent theoretical works, this limitation may be lifted by considering multi-resonator metasurfaces [23], [24]. In fact, in our devices the nonreciprocal ratios are mainly limited by the minimum transmission level, which also determines the transmission along the backward direction within the NRIR. As discussed above, in realistic devices the minima of the Fano lineshape remain above zero due to fabrication imperfections and spectral averaging induced by finite spectral and angular linewidths of our excitation/collection system. Improvement of these factors may therefore lead to a substantial increase of the observed nonreciprocal transmission ratio.

**Impact of Spectral Detuning on Intensity-Dependent Nonreciprocity.** In all measurements in Fig. 3, the excitation wavelength was kept close to the low-power transmission minima. This small detuning helps reducing the intensity required to observe a sizable change in transmission, but it also leads to a relatively smooth dependence of the transmission on the input intensity. However, as also suggested by the calculation in Fig. 1f, the same device can give rise to different intensity-dependent transmission curves as a function of the detuning between the impinging laser and the

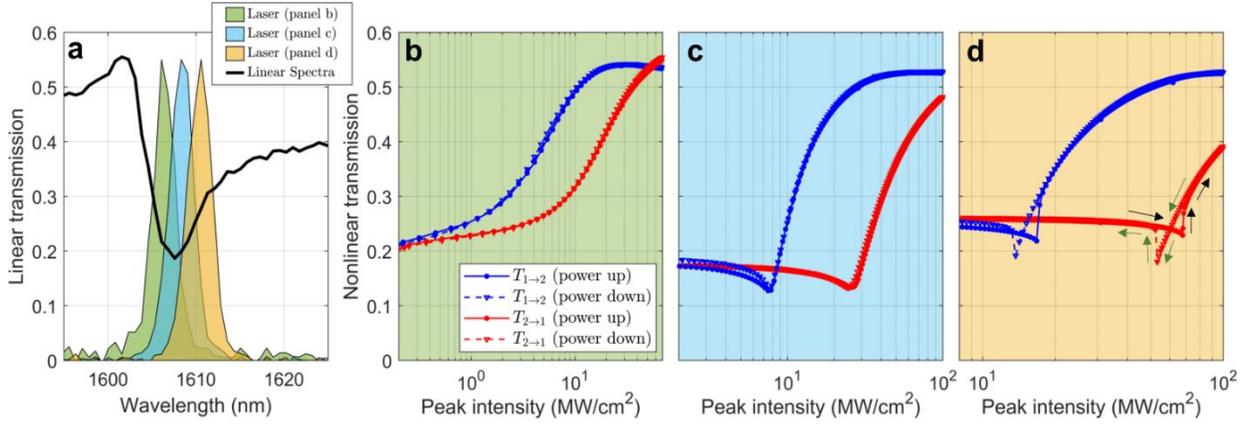

**Figure 4**. Nonreciprocal response of the same device for different excitation wavelengths. (a) Linear transmission spectra of the device under study (solid black lines), and three different excitation laser spectra (shaded areas, see legend). (b-d) Nonreciprocal response of the metasurface for the three different excitation wavelengths shown in panel a, as the intensity is swept up and down. Blue (red) symbols denote transmission under excitation from port 1 (2). The arrows in panel d denote the portions of the curve where the input intensity is increased (black arrows) or decreased (green arrows).

Fano transmission minima. Indeed, in Fig. 4 we report several measurements on the same device (whose linear transmission spectrum is shown in Fig. 4a, black curve) employing three different excitation laser spectra, shown by the three shaded areas in Fig. 4a. For each excitation, we measured the nonlinear transmission in the two directions [Figs. 4(b-d)], consistent with Fig. 3. All measurements show a clear nonreciprocal response, but with markedly different lineshapes: when the excitation laser is tuned very close to the steep edge of the Fano lineshape (green laser spectrum in Fig. 4a), a very smooth variation of the transmission curves is obtained at low intensities (Fig. 4b). As the excitation laser is progressively red-detuned (blue and yellow laser spectra in Fig. 4a), much sharper jumps occur in the intensity-dependent transmission curves (Figs. 4c and 4d), albeit at higher intensities. Such behavior can be useful to enhance the nonreciprocal transmission ratio at desired intensities, by aligning the transmission minima in the backward direction with transmission maxima in the forward direction.

**Bistability and Impact of Thermo-Optic Effects.** Importantly, at large detunings (Fig. 4d) the intensity-dependent transmission curves show a clear hysteresis behavior, whereby the values of transmission depend on whether the input intensity is increased or decreased, as pointed by the arrows in Fig. 4d. This phenomenon is expected in systems with third-order nonlinearities, and it is due to the existence of multiple stable steady-states for the same input power. Figure 5a shows a set of measurements (from a different device) where the bistability region is wider and clearer.

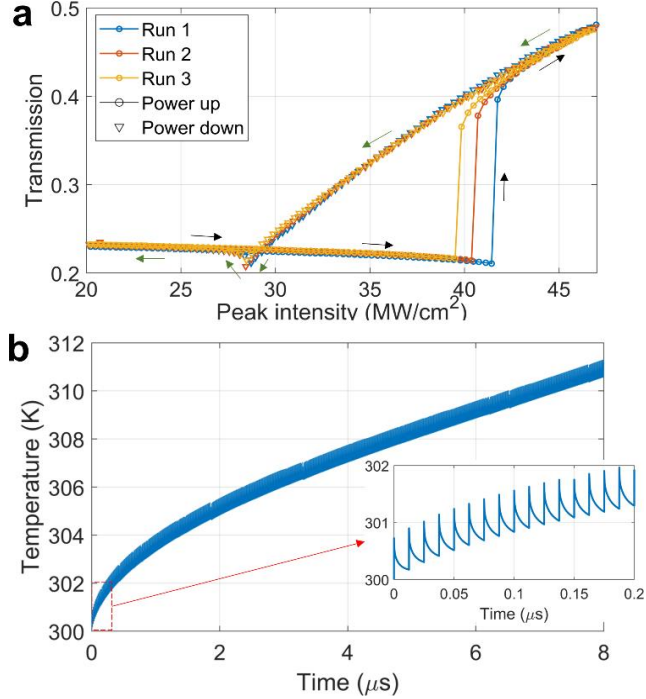

**Figure 5**. (a) Experimentally measured bistability response. We measured the forward transmission of the metasurface by scanning the power up and down three times. Different colors identify the three power ramps (see legend). Increasing (decreasing) powers are denoted by circles (downward triangles). (b) Numerical simulations of the time-dependent temperature variation at the center of the grating due to electromagnetic absorption and heating, assuming a pulsed excitation with repetition rate 80 MHz, a peak impinging intensity of 5 MW/cm$^2$ and absorption coefficient $\kappa$=0.01 for amorphous silicon (additional details in the SM).

Here we focus on the scenario in which the metasurface is excited from port 1 and, in order to verify the reproducibility of the bistability region, we repeated the up-and-down power ramp three times. For increasing intensities, the transmission experiences a sudden jump at a peak intensity of about $I_{peak} = 40$ MW/cm$^2$, corresponding to the transition from the first stable state to the second one. As the peak intensity is later decreased, the system remains in the second stable state until the peak intensity is about $I_{peak} = 27$ MW/cm$^2$, after which it jumps back to the first stable state.

The presence of a clear bistable regime provides an important hint to shed light on the origin of the nonlinearity at play in these experiments, and in particular on its characteristic timescale. In order to showcase bistability, a nonlinear optical resonator must be able to retain some memory of the previous history of its internal state. In resonators with *instantaneous* nonlinearities (i.e., where the intensity-dependent permittivity shift builds up and decays over times much shorter than any other relevant timescale), bistability can be observed when the power of a CW excitation is slowly swept up and down while maintaining the excitation uninterrupted. This adiabatic variation of the external excitation allows the system to populate different stable states depending on its previous state. In our experiment, while the time-averaged impinging intensity is slowly swept up and down (on timescales of tens of seconds), the instantaneous impinging intensity is quickly and repeatedly turned on and off due to the pulsed excitation, composed of short pulses of duration of ~2 ps separated by ~12 ns. Thus, any nonlinearity with a characteristic build-up time faster than few nanoseconds would not lead to a bistable region with clearly separated branches in our experimental conditions. This observation suggests that slower nonlinear effects, such a thermo-optic effects, may have a significant role in the observed nonlinearity-induced nonreciprocal response.

To corroborate the impact of thermo-optic effects, we numerically calculated the time-dependent temperature increase in our devices under realistic experimental conditions and material parameters (Fig. 5b) [35]. For simplicity, and since we are only interested in estimating the electromagnetic-induced temperature increase, in these simulations we assume that the refractive index of silicon is not affected by the temperature variation. Our numerical calculations (Fig. 5b) show that, for plane-wave illumination with uniform intensity of 5 MW/cm$^2$, the temperature at the center of the metasurface unit cell quickly rises by more than 10 degrees Celsius within a time span of about 10 microseconds. The inset in Fig. 5b reveals that this relatively large temperature increase is actually slowly built-up by each pulse, and therefore it highly depends on the specific value of the laser repetition rate. Assuming a thermo-optic coefficient $dn/dT = 2.3 \cdot 10^{-4} K^{-1}$ for amorphous silicon [36], a temperature increase of few tens of degrees will induce a relative spectral shift of the metasurface resonant frequency $\Delta\omega/\omega \sim 10^{-3} \div 10^{-2}$, which is large enough to account for the transmission changes observed in our experiments [35]. Thus, while fast nonlinearities (optical Kerr effect or carrier injection) may be playing a role in the observed nonreciprocity, our numerical simulations and the occurrence of a clear bistability region suggest that slower thermo-optic effects constitute the main nonlinearity at play in our experiments. As discussed above, thermal effects are enhanced here due to the high repetition rate (80 MHz) of the pulsed laser utilized. We expect that, by reducing the laser repetition rate and/or increasing the thermal dissipation of the metasurface, the influence of thermo-optic effects can be eliminated, and much faster nonlinearities may become dominant. On the other hand, slow nonlinearities have recently attracted much attention, and they have been shown to unlock novel phenomena such as non-Markovian dynamics [37], limit cycles and chaos [38], and nonreciprocal pulse shaping and chaotic modulation [39].

## Conclusions and Outlook

In this paper, we have experimentally demonstrated the occurrence of free-space fully-passive and bias-free nonreciprocity in tailored silicon metasurfaces leveraging nonlinear q-BICs. By combining the third-order nonlinearities of silicon with out-of-plane symmetry breaking we realized a metasurface for which the same input intensity beam, injected from opposite directions, leads to markedly different shifts of the refractive index, hence enabling large nonreciprocal responses for free-space illumination. By engineering the in-plane asymmetry of the metasurface, we have been able to accurately control the radiative linewidth of the metasurface and tailor the q-BIC, which allowed us to minimize the operating intensities while maintaining a large transmission contrast. We experimentally demonstrated nonreciprocal transmission over a large range of intensities, with nonreciprocal ratios larger than 10 dB and insertion loss lower than 3 dB for operating intensities smaller than 50 MW/cm$^2$. We further demonstrated that the nonreciprocal response can be largely tuned by simply controlling the thickness of a residual layer, which makes our design particularly simple and appealing for foundry-compatible fabrication. The values of insertion loss obtained in this work are close to the minimum value imposed by fundamental trade-offs, and further improvements could be obtained by considering multi-resonator devices [14], [15]. Our results demonstrate a powerful and broadly applicable route to obtain free-space fully-passive nonreciprocal propagation, by leveraging quasi-bound states in the continuum and material nonlinearities, which paves the way for several applications in, e.g., LiDAR, protection of high-power lasers, and nonreciprocal signal routing for analog and quantum computing.

**Acknowledgements**

The authors would like to thank Dr. Dmitriy Korobkin for experimental assistance in the earlier stages of this project. This work was supported by the Air Force Office of Scientific Research and the Simons Foundation. This work is part of the research program of The Netherlands Organization for Scientific Research (NWO).

**Author contributions statement**

All authors conceived the idea and the corresponding experiment. M.C., A.C. and D.S. designed the device and performed the numerical analysis. M.C. fabricated the devices and performed the experimental measurements with assistance from A.C.. All authors analyzed the data and contributed to writing the manuscript. A.A. and A. P. supervised the project.

**Competing Interests**

The authors declare no competing interests.

**Data availability**

The data that support the findings of this study are available from A.A and M.C. upon reasonable request.

# Passive Bias-Free Nonreciprocal Metasurfaces Based on Nonlinear Quasi-Bound States in the Continuum

Michele Cotrufo[1], Andrea Cordaro[2,3], Dimitrios L. Sounas[4], Albert Polman[3] and Andrea Alù[*1,5]

[1]*Photonics Initiative, Advanced Science Research Center, City University of New York, New York, NY 10031, USA*

[2]*Van der Waals-Zeeman Institute, Institute of Physics, University of Amsterdam Science Park 904, 1098 XH Amsterdam, The Netherlands*

[3]*Center for Nanophotonics, AMOLF, Science Park 104, 1098 XG Amsterdam, The Netherlands*

[4]*Department of Electrical and Computer Engineering, Wayne State University, Detroit, Michigan 48202, USA*

[5]*Physics Program, Graduate Center of the City University of New York, New York, NY 10016, USA*

## S.1 Sample Fabrication

The samples were fabricated with a standard top-down lithographic process. Glass coverslides (25 x 75 x 1 mm, Fisher Scientific) were used as transparent substrates. The substrates were cleaned by placing them in an acetone bath inside an ultrasonic cleaner, and later in an oxygen-based cleaning plasma (PVA Tepla IoN 40). After cleaning, a layer of approximately 100 nm of amorphous silicon (α-Si) was deposited via a plasma-enhanced chemical vapor deposition (PECVD) process. A layer of e-beam resist (ZEP 520-A) was then spin-coated on top of the samples, and the desired pattern was written with an electron beam tool (Elionix 50 keV). After ZEP development, the pattern was transferred to the underlying silicon layer via dry etching in an ICP machine (Oxford PlasmaPro System 100). Different copies of the same sample were fabricated in the same EBL run, and then etched with different etching times in order to control the residual thickness $t$ (see design in Fig. 1). The resist mask was finally removed with a solvent (Remover PG).

## S.2 Optical Characterization

The linear and nonlinear responses of the samples were measured with a custom-built setup shown in Fig. S1. To measure the linear transmission of the gratings, a collimated broadband light (Thorlabs, SLS201L) was linearly polarized and then weakly focused on the sample with a lens (L1) with f = 20 cm focal length. The transmitted signal was collected on the other side of the sample by an identical lens, and redirected to a spectrometer (Ocean Optics). The transmission spectra were then obtained by measuring the spectra of the lamp with and without the sample, and normalizing the former by the latter. For the nonlinear measurements, an optical parametric oscillator (APE, Levante IR ps) was used to generate a tunable pulsed laser (pulse duration $\tau = 2$ ps, repetition rate f = 80 MHz). The linewidth of the laser in the spectral region of interest (1400 nm - 1600 nm) is ~2-3 nm. The polarization and power of the laser were controlled by cascading a half waveplate (HWP) and a linear polarizer (LP), with the linear polarizer oriented along the direction of the grating wires (i.e. y-direction in Fig. 1b of the main text). The laser beam was weakly focused on the center of the metasurface under test, with the same lens L1 used in the linear measurement. The transmitted signal is collected by a second identical lens (L2), and redirected (via flip mirrors) either to the photodector P1, or to the spectrometer (to check the spectral content of the laser) or to a NIR camera (Xenics Xeva 320) for alignment purposes.

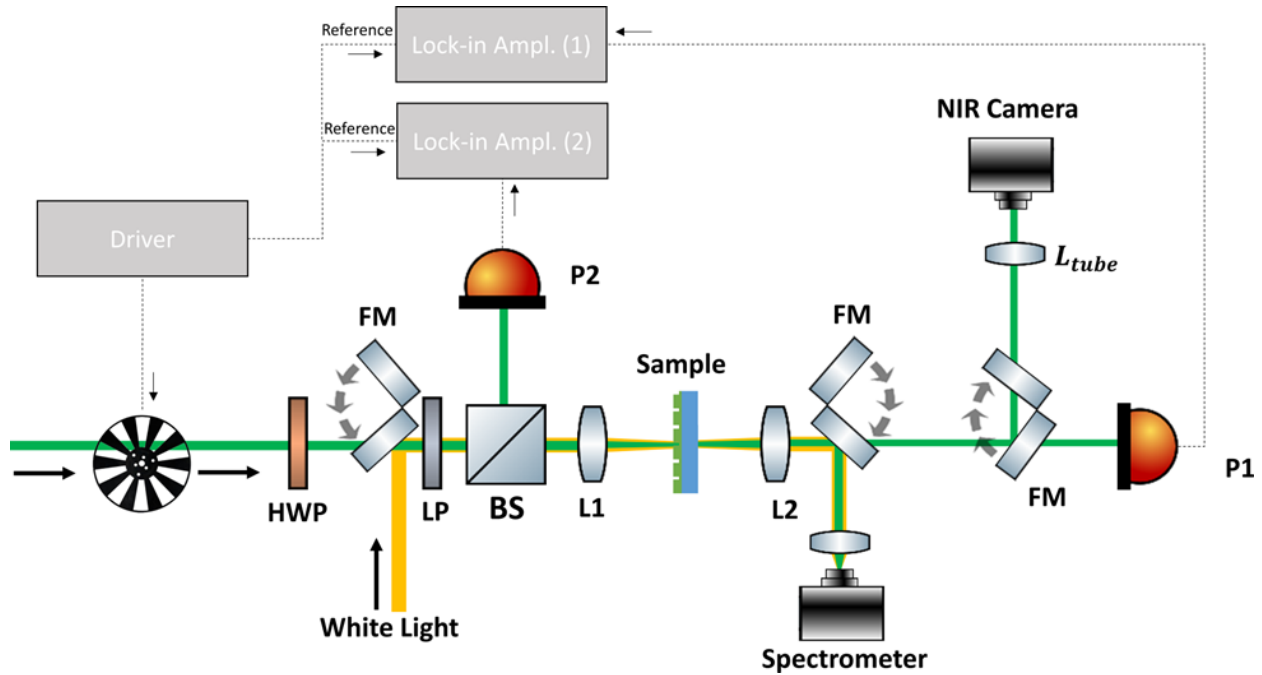

**Figure S1**. Schematic of the experimental setup (see text for details).

Two identical photodiodes P1 and P2 (Thorlabs, DET50B2) were used to measure the transmission level through the metasurface. A beamsplitter (BS) placed before the excitation lens L1 was used to redirect approximately 10% of the laser power to the photodiode P2, for reference. Each photodiode is connected to a separate lock-in amplifier (Stanford Research Systems, models SR810 and SR865A), and the input laser was mechanically chopped with a modulation rate of 500 Hz. To acquire the power-dependent transmission curves shown in Figs. 3-5 of the main paper, the HWP was mounted on a motorized rotation stage (Thorlabs, K10CR1). A custom-built software was used to slowly vary the impinging power by rotating the HWP while simultaneously recording the signal measured by the two lock-in amplifiers. In each measurement run, the impinging power was increased up to the maximum value displayed in each plot in Figs. 3-5 and then decreased, without interruption. The sample was then flipped, and the procedure was repeated to measure the power-dependent transmission along the opposite direction.

To obtain the absolute transmission values, the same experimental procedure was repeated with and without the metasurface, and the latter measurement was used for normalization. The average input power $P_{\text{avg}}^{(meas)}$ was measured by an additional calibration run performed with a thermal powermeter (Thorlabs, S401C) placed before the excitation lens L1. We note that the measured $P_{\text{avg}}^{(meas)}$ is affected by the duty cycle of the chopper (50%), and we therefore define $P_{avg} = 2 \times P_{\text{avg}}^{(meas)}$. To retrieve the impinging peak intensities, we measured the radius of the laser spot on the metasurface by imaging both the spot and a whole metasurface field with the same imaging lenses, and by using the known length of the metasurface field (700 µm) as a ruler. The spot radius $w_0$ (defined as the distance from the center at which the intensity drops by a factor $e^2$) was determined independently before each measurement run, with typical values lying in the range $w_0 \approx 80 \div 90$ µm. Following standard formulas, the impinging peak intensity (both in time and space) was then calculated by

$$I_{peak} = \frac{2 P_{avg}}{f\,\tau\,\pi w_0^2} \quad \text{(S1)}$$

where $f$ and $\tau$ are the repetition rate and the pulse duration of the laser, respectively.

## S.3 Experimental protocol to avoid spurious effects

The central experimental result discussed in this work is that a beam with the same wavelength and power experiences markedly different transmission levels when impinging on the *same* metasurface from different sides. In our experiment, this is obtained by first measuring the power-dependent transmission through the metasurface when the laser impinges from port 1, and then physically flipping the sample and repeating the measurement. In order to make sure that the observed effects are truly due to nonreciprocity it is very important to make sure that, in the two set of measurements (i.e. excitation from port 1 and from port 2), the excitation configurations are exactly the same. Indeed, the linear transmission spectra along the two directions might be quite different if the two measurements are done, for example, at slightly different impinging angles (due to the linear dispersion of the lattice modes), or a different points of the metasurface (due to fabrication-induced disorder). To minimize the influence of these effects, the following protocol was used in the experiments prior to measure the power-dependent transmission along each direction:

1. To ensure that the input beam impinges at normal incidence, we monitored the back-reflected spot from the sample on a pinhole placed right after BS in Fig. S1. By making sure that the back-reflected spot is aligned with the path of the impinging laser, we ensured that the metasurface is normal to the impinging beam with a maximum error of $\Delta\theta < 0.5°$.
2. The in-plane orientation of the metasurface was checked by imaging it with the NIR camera shown in Fig. S1. The magnification of our imaging system resulted in approximately 1 pixel = 10 µm on the camera, which allows us to estimate a maximum error of $\Delta\phi < 0.8°$ in the azimuthal angular position of the metasurface.
3. To ensure that the metasurface layer is at the same transversal position along the optical axis in the two measurements, we adjusted the transversal position of the sample until the metasurface edges were well-focused in the camera. The radius of the laser spot at the metasurface position was then measured with the method described above. In each measurement run we verified that the radius of the two spots (for the two excitation directions configurations) were identical within a maximum discrepancy of ~2%, in line with our imaging resolution. Moreover, we note that our experiment is not expected to be particularly sensitive to the exact transversal position of the metasurface: for the excitation configuration considered here (excitation focal length = 200 mm, spot diameter before the lens ≈ 4 mm, wavelength ~ 1550 nm), the depth of focus is expected to be approximately 10 mm, much larger than the total thickness of our sample (~ 1mm).

4. Finally, linear transmission spectra were acquired immediately before each nonlinear measurement *from the same area of the metasurface where the nonlinear measurement was performed*. The linear transmission spectra were acquired again after each nonlinear measurement, to confirm that (1) no substantial drift of the sample/setup occurred and that (2) no irreversible damage occurred to the sample. In each panel of Fig. 3 of the main text, the left-side plots show the transmission spectra acquired right before the nonlinear measurements along each direction, as blue and red curves.

## S.4 Electromagnetic asymmetry due to the substrate and the slanted sidewalls

The measurements in Fig. 3e show that a device without any residual silicon layer (t=0) can still provide some nonreciprocal transmission, albeit in a much smaller range of intensities (NRIR = 1.45). As explained in the main text, the nonreciprocal intensity range is intimately connected to the electromagnetic asymmetry in the linear regime, that is, the ratio $\kappa = |E_1|^2/|E_2|^2$ between the field intensities $|E_i|^2$ ($i = 1,2$) induced at the same point when the same power is injected from either port 1 or port 2. In particular, NRIR = $\kappa$. In our devices we varied the residual layer thickness t in order to control the electromagnetic asymmetry, and thus the NRIR. However, even for t=0, the vertical symmetry of the metasurface is still broken by two factors, namely (1) the asymmetry between the substrate (glass) and superstrate (air), and (2) the possible presence of slanted sidewalls. Both effects can induce an additional electromagnetic symmetry. To evaluate these effects, we performed numerical calculations (Comsol) of a device with t=0 and with slanted sidewalls, quantified by the angle $\theta$ in Fig. S2a. The other parameters are similar to the devices considered in Figs. 1-3 of the main text. For each value of $\theta$ we calculated the electric field $|E_i|^2$ generated at the same point (star symbol in Fig. S2a) when the same plane wave is injected from one of the two ports. As shown in Fig. S2b, the electromagnetic asymmetry $|E_1|^2/|E_2|^2$ can be quite large even for moderately slanted sidewalls. Importantly, even for perfectly vertical sidewalls ($\theta = 90°$) the device retains a non-negligible electromagnetic asymmetry due to the presence of the substrate. According to these simulations, the value of NRIR=1.45 observed in our experiment for t = 0 (Fig. 3e in the main text) can be explained by a value of $\theta \approx 80°$, which is in line with the typical values obtained in fabrication.

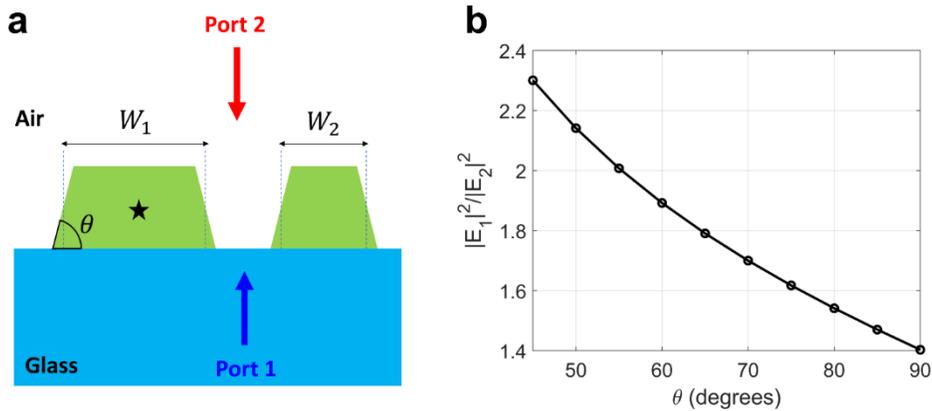

**Figure S2**. (a) Schematic of the simulated design. We consider the general case in which the sidewalls are not necessarily vertical, as quantified by the angle θ. (b) Calculated electromagnetic asymmetry $|E_1|^2/|E_2|^2$ between the field intensities $|E_i|^2$ ($i = 1,2$) induced at the same point (star symbol in panel a) when the same power is injected from either port 1 or port 2, versus the angle θ.

## S.5 Heat transfer numerical simulations

In order to estimate the influence of thermo-optic effects we performed numerical simulations by solving the heat transfer and Maxwell equations using Comsol Multiphysics. To reduce the numerical cost of simulations, we performed 2D simulations assuming an infinitely extended metasurface and a plane-wave excitation. Figure S3 shows the simulated geometry. We considered a representative geometry with lattice constant a = 750nm, gaps G = 100 nm, total thickness H = 100 nm, residual thickness t = 20 nm, $W_1$ = 300 nm and $W_2$ = 250 nm. Following ref. [1], we assumed that the imaginary part of the refractive index of α-Si is $\kappa = 0.01$.

The electromagnetic excitation is a plane wave linearly polarized along the y-direction, propagating towards the positive z-direction, and with a frequency tuned to the transmission minimum of the Fano lineshape of the considered device. The plane-wave amplitude is time-dependent and described by $E(t) = E_0 \cdot \sum_{k=0}^{N} p(t - k/f)$, where $f$ = 80 MHz is the repetition rate of the laser, and $p(t) = \exp(-1.38 \cdot t^2 / \tau^2)$ is a Gaussian pulse with duration $\tau$ = 2 ps. The amplitude $E_0$ is adjusted in order to simulate a spatially invariant impinging intensity of 5 MW/cm$^2$. By solving Maxwell equations, we calculated the induced electromagnetic field $\mathbf{E}(\mathbf{r}, t)$ (Fig. S2a), and the time- and space-dependent absorption profile (Fig. S2b)

$$S(\mathbf{r},t) = \frac{1}{2}\omega\varepsilon_0 \,|\,\mathrm{Im}(\varepsilon_{Si})\,|\cdot|\,\mathbf{E}(\mathbf{r},t)\,|^2 \tag{S2}$$

where $\omega$ is the excitation frequency and $\varepsilon_{Si}$ is the relative permittivity of Silicon. The absorbed power density $S(\mathbf{r},t)$ is then used a heat source in the heat transfer equation

$$\rho C_p \frac{\partial T}{\partial t} + \nabla \cdot (-\kappa_{th}\nabla T) = S \tag{S3}$$

where $T$ is the temperature, $\rho$ is the material density, $C_p$ is heat capacity at constant pressure and $\kappa_{th}$ is the thermal conductivity. The values of these parameters used in the simulations are reported in table S1. We used periodic boundary conditions on the lateral boundaries (i.e. the boundaries orthogonal to the x-direction in Fig. S3a), while for the top and bottom boundaries we assumed that the structure can exchange heat with the environment via convection and radiation. Convection is simulated by adding an outgoing convective heat flux term to eq. S3 of the form $q_{conv} = h\ (T_{ext} - T)$, where $T_{ext} = 300$ K and $h$ is the heat transfer coefficient (assumed $h = 1.4$ W/(m·K) for the glass interface and $h = 5$ W/(m·K) for the air interface). Radiation is simulated by adding an outgoing heat flux $q_{rad} = \varepsilon\sigma(T_{ext}^4 - T^4)$, where $\varepsilon$ is material emissivity ($\varepsilon = 0.8$ for air and $\varepsilon = 0.95$ for glass [2]) and $\sigma$ is the Stefan–Boltzmann constant.

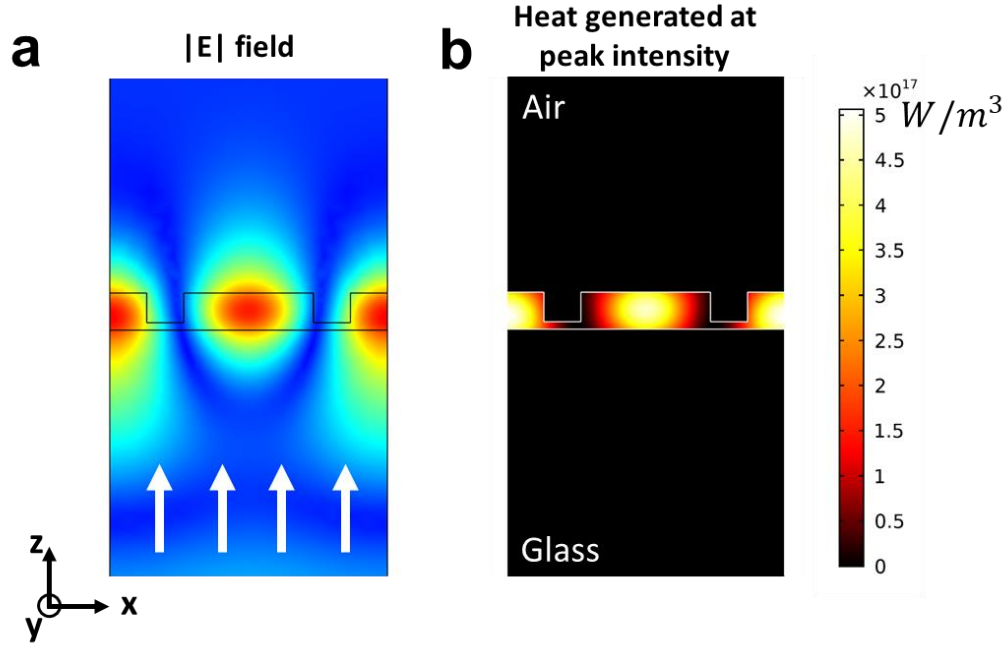

**Figure S3**. (a) Electric field generated in the structure upon plane wave illumination. (b) Electromagnetic heating (corresponding to the absorbed power density defined in eq. S2) generated when the impinging plane wave has a spatially uniform intensity of 5 MW/cm$^2$.

In order to reduce the computational cost, in these simulations we did not explicitly assume that the refractive index of silicon depends on the local temperature. Instead, we assumed that the refractive index is constant, and we used the calculated temperature variation $\Delta T$ (see Fig. 5 in the main text), together with standard values of the thermo-optic coefficient of $dn/dT = 2.3 \cdot 10^{-4} K^{-1}$ for amorphous silicon [3], to estimate the refractive index variation $\Delta n \approx dn/dT \times \Delta T$. By using standard first-order-approximation formulas (see, e.g., eqs. 28-29 in ref. [4]), we can estimate the relative shift of the resonant wavelength, $\Delta\lambda/\lambda \approx -\Delta n/n$, induced by a certain temperature variation. The calculate temperature increase of $\Delta T \approx 10$ K after 8 $\mu s$ (see Fig. 5b of the main text) translates into a relative shift $\Delta\lambda \approx 1$ nm (assuming $\lambda = 1550$ nm). Assuming that the temperature variation versus time remains initially linear, we expect the spectral shift to be of the order of $\Delta\lambda \approx 5 \div 10$ nm within hundreds of microseconds. This detuning is large enough to account for the power-dependent transmission changes observed in this work. We note that, in a more realistic setting, as the refractive index (and thus the resonant wavelength) is shifted, the coupling efficiency of the external pump to the electromagnetic mode will eventually decrease, thus introducing a negative feedback that limits the temperature increase.

| **Material / Property** | $\rho$ [Kg·m$^{-3}$] | $C_p$ (J·Kg$^{-1}$·K$^{-1}$) | $\kappa_{th}$ (W·m$^{-1}$·K$^{-1}$) |
|---|---|---|---|
| Silicon | 2330 | 800 | 1.8 |
| Glass | 2210 | 730 | 1.4 |
| Air | 1.17 | 1005 | 0.026 |

Table S1.